# Controlling magnetism with light in a zero orbital angular momentum antiferromagnet

M. Matthiesen[1], J. R. Hortensius[1], S. Mañas-Valero[2], M. Šiškins[1], B. A. Ivanov[3,4], H. S. J. van der Zant[1], E. Coronado[2], D. Afanasiev[4], A. D. Caviglia[1,5]

[1]*Kavli Institute of Nanoscience, Delft University of Technology, P.O. Box 5046, 2600 GA Delft, Netherlands*

[2]*Instituto de Ciencia Molecular (ICMol), Universitat de Valencia, Catedrático José Beltrán 2, 46980 Paterna, Spain.*

[3]*Institute of Magnetism, National Academy of Sciences and Ministry of Education and Science, 03142 Kyiv, Ukraine*

[4]*Radboud University, Institute for Molecules and Materials, 6525 AJ Nijmegen, Netherlands*

[5]*DQMP–University of Geneva, École de Physique, 24, Quai Ernest-Ansermet, CH-1211 Geneva, Switzerland*

Antiferromagnetic materials feature intrinsic ultrafast spin dynamics, making them ideal candidates for future magnonic devices operating at THz frequencies. A major focus of current research is the investigation of optical methods for the efficient generation of coherent magnons in antiferromagnetic insulators. In magnetic lattices endowed with orbital angular momentum, spin-orbit coupling enables spin dynamics through the resonant excitation of low-energy electric dipoles such as phonons and orbital resonances which interact with spins. However, in magnetic systems with zero orbital angular momentum, microscopic pathways for the resonant and low-energy optical excitation of coherent spin dynamics are lacking. Here, we consider experimentally the relative merits of electronic and vibrational excitations for the optical control of zero orbital angular momentum magnets, focusing on a limit case: the antiferromagnet manganese thiophosphate ($MnPS_3$), constituted by orbital singlet $Mn^{2+}$ ions. We study the correlation of spins with two types of excitations within its band gap: a bound electron orbital excitation from the singlet orbital ground state of $Mn^{2+}$ into an orbital triplet state, which causes coherent spin precession, and a vibrational excitation of the crystal field that causes thermal spin disorder. Our findings cast orbital transitions as key targets for magnetic control in insulators constituted by magnetic centers of zero orbital angular momentum.

Magnets are central to modern advanced technologies, for example, as bits in magnetic memories or as spin current conductors in spintronic devices. Recently, the inclusion of antiferromagnets into spin-based devices has garnered much interest, as they offer faster intrinsic spin dynamics and less energy dissipation than the more commonly utilized ferromagnets [1]. Lacking net magnetization, antiferromagnets must be manipulated through non-magnetic drives, such as electric currents [2], voltage [3], mechanical strain [4,5], or light [6]. The latter, in the form of ultrashort laser pulses, allows spin perturbation at diabatic timescales, beating the intrinsic timescales even of antiferromagnetic spin dynamics [7]. The light-spin coupling typically involves excited states of the magnetic ion orbitals, either virtually [8] or directly [9,10], by causing a transient change of orbital angular momentum. This modifies the spin-lattice coupling (e.g. through magnetocrystalline anisotropy), and exerts a transient torque on the spins. With the advent of intense low-frequency pulses capable of resonant excitation of optical phonons, new approaches have become possible that target the spin-lattice coupling not through orbitals, but instead through perturbation of the crystal field [11]—indirectly causing dynamics of the orbital angular momentum.

The magnetic ions of many antiferromagnets lack orbital angular momentum, either due to strong crystal field quenching (e.g. $Ni^{2+}$ in an octahedral field), or because the ion has a half-filled shell (e.g. $Mn^{2+}$, $Fe^{3+}$, $Ni^{3+}$). This raises the question of whether, in such magnets, the most efficient pathway to magnetic control should be via lattice or orbital perturbation. We study the magnetic dynamics caused by excited states of the crystal lattice and of the electron orbits in manganese thiophosphate ($MnPS_3$). This two-dimensional antiferromagnet has recently attracted a lot of attention with observations of long-distance magnon transport [12], relativistic domain-wall dynamics [13] and giant modulations of its optical nonlinearities by Floquet engineering [14]. Notably, this material has zero orbital angular momentum due to the half-filled shell of the spin-carrying $Mn^{2+}$ ions, which are in an orbital singlet ground state. Perturbative sources of momentum are also minimal due to the very large energy difference (1.9 eV) between the ground state and the first excited state of $Mn^{2+}$. Additionally, the ligands are all light *p*-orbital ions and should contribute no substantial momentum. Altogether, this casts $MnPS_3$ as an interesting limit-case "orbital moment-free" antiferromagnet. We find that in this material, optical excitation of bound electrons involving a change of orbital moment can cause coherent spin precession, whereas excitation of lattice vibrations leads to a thermal spin disorder.

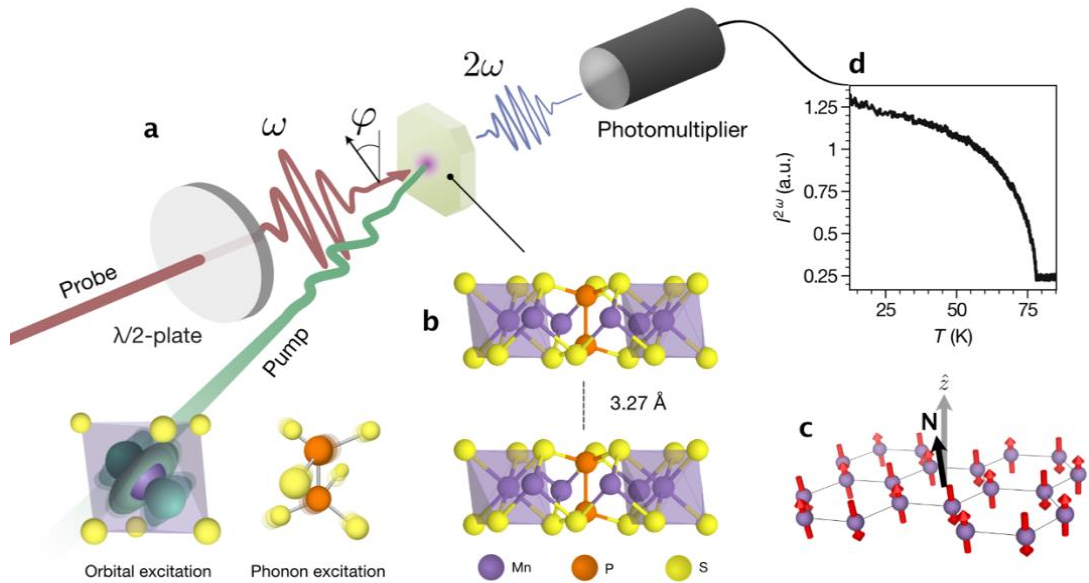

Figure 1 **Ultrafast optical spectroscopy of the magnetic order in single crystal MnPS$_3$. a**. Schematic of the all-optical experimental setup, consisting of two incident beams: One to excite the material (green) and one to generate the second harmonic probe (red, frequency *ω*). The generated SHG (2*ω*) is detected with a photomultiplier tube. *Inset*: To bring the system out of equilibrium, we excite either *d-d* orbital transitions of the magnetic ions or in the phonon band. **b.** The layered crystal structure, consisting of sulfur-coordinated $Mn^{2+}$ ions conjoined by $P_2S_6^{4-}$ molecules. **c.** The magnetic structure of MnPS$_3$, with **N** indicating the direction of the Néel vector and *z* the normal to the crystal planes. **d.** The temperature dependence of the second-harmonic intensity.

We study a single crystal sample of MnPS$_3$ grown by chemical vapor transport (details in ref. [15]), approximately 10 µm thick (see Supplemental Material S1) with a diameter of about 5 mm. This is a van der Waals crystal with a layered two-dimensional crystalline and magnetic structure [16] (see Fig. 1b,c), which due to its small interplane magnetic exchange [17] retains magnetic order down to a few layers [18,19]. Within the two-dimensional layers, the manganese Mn$^{2+}$ ions are arranged in a honeycomb lattice. Below a Néel temperature of $T_N$ = 78 K a fully compensated antiferromagnetic Néel ordered ground state forms (see Fig. 1c). This results in two collinear magnetic sublattices with oppositely oriented magnetizations, and their difference constitutes the antiferromagnetic order parameter—the Néel vector ***N***. A predominantly out-of-plane orientation of ***N*** is caused by the magnetic dipole-dipole interaction [20]. Despite the absence of magnetization, the magnitude of this order parameter can be measured optically: The magnetic point-group of MnPS$_3$ breaks space inversion symmetry, which allows for magnetic second harmonic generation (MSHG) [18,21,22]. This amounts to emission of light at frequency $2\omega$ in response to excitation at $\omega$, with the emission intensity $I^{2\omega}$ being quadratic in $N$ [14]. Since the crystal lattice itself has a center of inversion, there is no crystalline contribution to second harmonic generation in the dipole approximation. A small electric quadrupole contribution is present, but is more than one order of magnitude smaller than the spin-induced electric dipole emission causing MSHG. We confirm the sensitivity of SHG to the magnetic order parameter by measuring its temperature dependence (Fig. 1d) which shows a sharp onset of signal at a temperature of $T$ = 78 K (see Supplemental Material S2 for further details).

Below the Mott band gap (2.96 eV) [23], MnPS$_3$ has several absorption lines of different origin. We first consider the absorption lines caused by transitions within the Mn$^{2+}$ ions. The ionic ground state $^6A_{1g}$ ($t_{2g}^3 e_g^2$) is an orbital singlet ($L$ = 0) with five unpaired spins ($S$ = 5/2). Due to the S$^{2-}$ crystal field, the first excited state is split into four states, with the lowest $^6T_{1g}$ ($t_{2g}^4 e_g^1$) being an orbital triplet with $S$ = 3/2 at an energy of 1.92 eV, well below the band gap [23]. Since this transition changes the quantum numbers of the magnetic centers, we find it a promising target for optical perturbation of the magnetic order. This state can in fact be populated optically: While prohibited by both spin- and parity-conserving selection rules, this transition does carry a dipole moment due to coupling to locally symmetry-breaking optical phonons [24].

To excite this orbital transition, we generate pulses with photon energies in resonance with the transition energy, using an optical parametric amplifier (TOPAS, Light Conversion) seeded by an amplified Ti:Sapphire laser (Coherent Astrella, $\Delta t$ = 100 fs, $f_{rep}$ = 1 kHz). We measure the ultrafast light-induced magnetic dynamics of the material in a stroboscopic pump-probe experiment, using MSHG as the magnetic probe (see Fig. 1a).

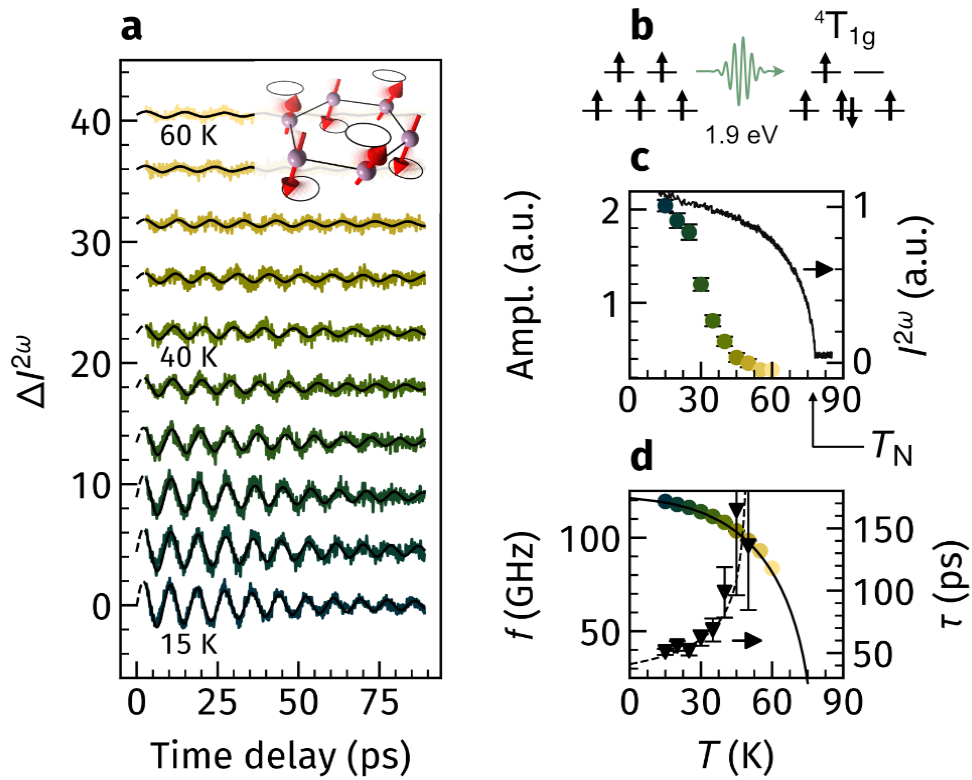

*Figure 2* **Excitation of a magnon through orbital transitions. a.** Pump-induced changes in the MSHG intensity in response to excitation with photon energy $\hbar\omega$ = 1.9 eV, for temperatures 15–60 K. Non-oscillatory background signal is subtracted. Raw traces and Fourier transforms are available in Supplemental Material S3. **b.** Single-particle Mulliken notation view of the optical transition excited. **c.** Extracted oscillation amplitude (in green-yellow, left axis), obtained through best-fit to a damped sine wave in panel a. The amplitude is normalized on the total MSHG intensity (black, right axis). Its dependence on pump fluence and probe polarization is given in Supplemental Material S4 and S5, respectively. **d.** Best-fit oscillation frequency (green, left axis) for each temperature, and the damping time $\tau$ (black, right axis), with a dotted line as guide to the eye.

Figure 2a shows the intensity modulation of the MSHG in response to excitation into the $^4T_{1g}$ state (see Fig. 2b), using an incident polarization at which the intensity was maximal. The time-domain signals reveal oscillatory dynamics at a frequency of 119 GHz (0.49 meV), which coincides with the magnon energy gap as measured by neutron scattering [16]. This magnon mode involves the precession of antiferromagnetically oriented spins causing a deviation of the Néel vector **N** from its equilibrium orientation. The initial phase of the resulting oscillations is close to zero, indicating a predominantly impulsive driving force imparted by the short-lived excited state. Approaching the magnetic phase transition from low temperatures (see Fig. 2d), we observe the mode slowing down along with a reduction of the antiferromagnetic order parameter *N*, as is typical for soft modes in the vicinity of phase transitions [25]. Surprisingly, the amplitude of the magnon decays at a temperature rate much faster than both the frequency and the magnetic order itself, and vanishes above approximately 50 K, significantly lower than $T_N$ = 78 K (Fig. 2c). Furthermore, the oscillation lifetime markedly increases as we approach this same sub-$T_N$ temperature (Fig. 2d), a highly unusual behavior for magnons. These temperature dependencies can arise either due to changes in the cross-section of the detection or excitation process, or in the dynamical properties of the material itself. In support of the latter, sub-$T_N$ features have been observed in other settings, such as Raman scattering [26], nanomechanical resonators [15] and neutron scattering [27,28].

To confirm the resonant character of the magnon excitation, we vary the photon energy of the excitation pulse across the $^4T_{1g}$ resonance and monitor the amplitude of the magnon mode. We observe a magnon amplitude only when the photon energy lies within the $^4T_{1g}$ absorption line, whereas indications of magnetic dynamics are not observed when exciting at energies below the orbital resonance (0.95 eV and 0.89 eV), as shown in Fig. 3. The immediate and phase-coherent spin dynamics launched via the excitation of the orbital resonance points to a direct impact of the excitation on the spins. We note that the phase of the spin precession is independent of the polarization state of the excitation pulse, including for circularly polarized pulses.

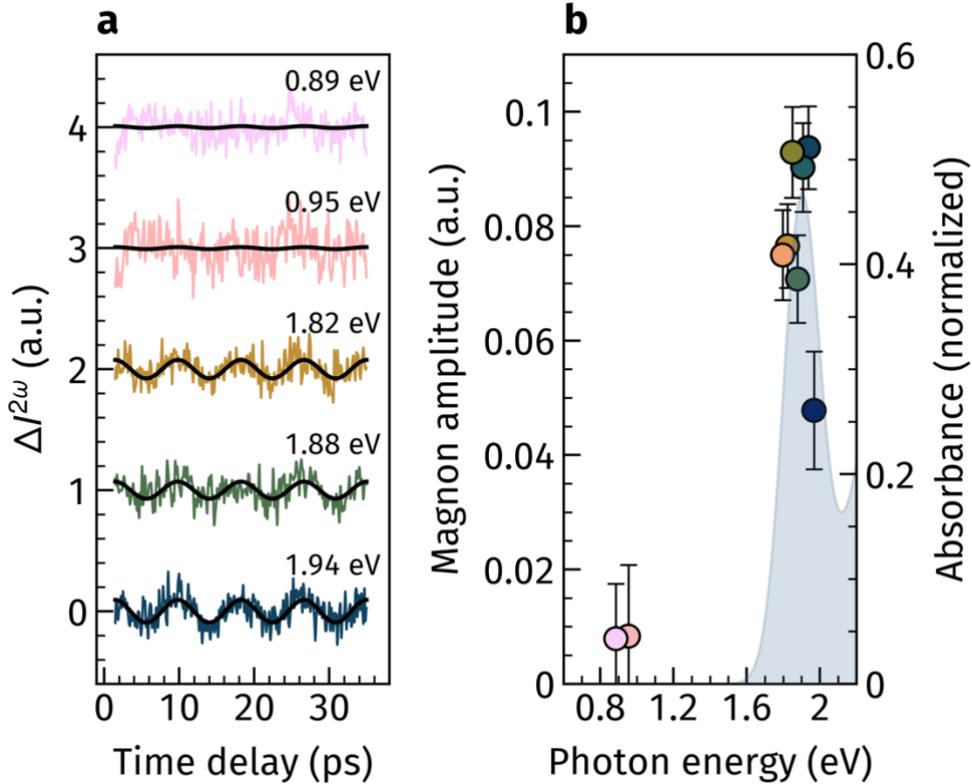

*Figure 3* **Pump photon energy dependence of the magnon amplitude. a.** Time-traces of the MSHG intensity changes in response to excitation with pump pulses of central photon energies in the range 0.88—1.97 eV. A polynomial background has been removed from the raw data, and the traces are offset. **b.** Extracted oscillation amplitude obtained through best-fit to a sine wave with $f$ = 119 GHz and fixed phase. The magnon is only excited in the vicinity of the $^4T_{1g}$ excited state, whose absorbance (Grasso, Phys. Rev. B 1991 **44** 20) is shaded in the background, normalized over the sub-gap *d-d* energies.

Our observation that an excitation pulse in resonance with the $^4T_{1g}$ orbital state can cause spin precession agrees well with demonstrations of selective optical control of magnetism via optical excitation of *d*–shell transitions in other antiferromagnets [10,29,30]. The spins in MnPS$_3$, while largely oriented out of plane, are rotated by a few degrees into the plane due to a small easy-plane anisotropy [31]. The excitation of electrons into $^4T_{1g}$, which carries orbital angular momentum, will enable a transient coupling of spin and orbital angular momentum. This, we argue, is the cause of spin precession: Since the spins are misaligned with respect to the crystallographic axes, and consequently the orbitals, the sudden coupling of spin and orbital angular momentum reorients the magnetic anisotropy direction throughout the $^4T_{1g}$ lifetime. This amounts to an impulsive torque on the spin, and causes spin precession, akin to previous observations in NiPS$_3$ [10].

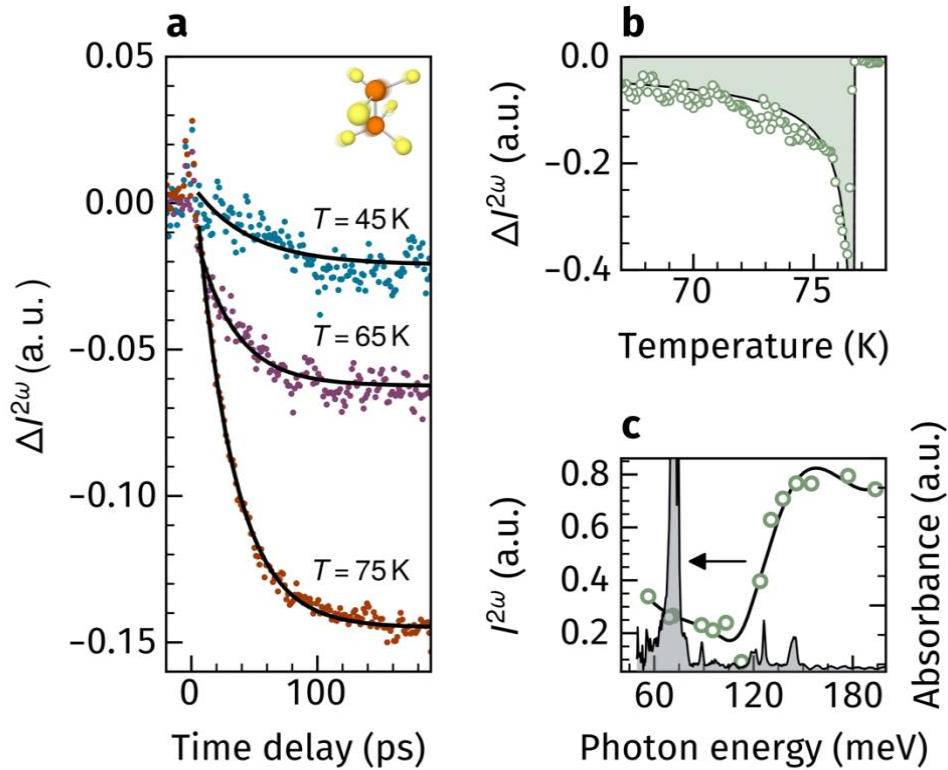

*Figure 4* **Magnetic quenching by phonon excitation. a.** Time trace of the magnetic dynamics induced by a pump of photon energy 120 meV (for which the quenching is maximal), at three different temperatures. Black lines are guides to the eye. The inset illustrates the local dynamics of the phonon. **b.** The change in MSHG measured at the fixed time delay of 0.6 ns, as a function of temperature, and exciting at 120 meV. **c.** The MSHG signal 0.6 ns after excitation, measured as a function of pump photon energy at 75 K. The sample absorbance as measured with FTIR is shown in gray.

Having demonstrated that excitation of an orbital transition leads to high-frequency phase-coherent spin dynamics, we turn to studying the effect of resonant lattice excitation on the magnetic state of MnPS$_3$. For the excitation of optical phonons, whose energies are much lower than the orbital transitions, we generate intense low-energy pulses: Using two pulses produced by a pair of optical parametric amplifiers, we perform difference frequency mixing in a 350 μm thick GaSe crystal. The highest-energy phonon modes, of $A_u$ and $B_u$ symmetry (75 meV), involve atomic motions within the ethane-like $(P_2S_6)^{4-}$ complexes [32]. Importantly, these modes do not involve motion of the $Mn^{2+}$ ions, and therefore preserves the distance between the adjacent $Mn^{2+}$ ions, leaving their magnetic dipole-dipole interaction unperturbed to first order. Qualitatively, the coupling of phonons to spins should then occur only via the crystal field experienced by individual $Mn^{2+}$ ions.

Figure 4a shows the change in MSHG, $\Delta I^{2\omega}$, in response to long-wavelength excitation, for three different temperatures. Close to the transition temperature, we observe a quenching of the MSHG signal, with an exponential decay of about 30 ps. Note that this quenching significantly exceeds the period of coherent spin precession $1/f_m$ = 8.4 ps, signifying thermally driven and phase-incoherent spin dynamics. Furthermore, we find that the signal does not recover between laser pulses, meaning the quenched state lasts for at least one millisecond (just before the next pulse arrival). At lower temperatures however, this quenching vanishes almost completely. To capture a detailed temperature dependence of the quenching magnitude, we fix the pump-probe delay late in the evolution (at 0.6 ns), and track the pump-induced signal ($\Delta I^{2\omega}$) as the sample is cooled down (see Fig. 4b). The quenching approaches zero at low temperatures, and rises sharply close to the Néel transition. This observation is a clear indication that although the lattice excitation quenches the Néel order in $MnPS_3$, these changes are thermal in nature. Next, we vary the pump photon energy, and track the total intensity ($I^{2\omega}$) of the MSHG at a fixed pump-probe delay of 0.6 ns, shown in Figure 4c. Sufficiently close to the phonon band, there is large magnetic quenching, while at higher energies this does not occur. The dynamics are therefore likely mediated by the population of optical phonons. Note that due to saturation effects and penetration depth mismatch between the short-wavelength probe pulses and long-wavelength excitation pulses, the spectral dependence of the quenching does not simply follow the phonon absorption line, as similarly observed in e.g. Ref. [33]. The slow timescale of the quenching dynamics, together with the dependence on the pump photon energy, supports the thermal origin of the magnetic quenching.

Optical excitation of the crystal field environment, via optical phonons, leads to slow thermal spin dynamics in $MnPS_3$. These dynamics are not phase coherent, and in this sense optical phonons do not offer themselves as a tool for magnetic control in this system. We argue that because of the sphericity of the $Mn^{2+}$ electrons, there is no orbital momentum to mediate a correlation between atomic displacements and spin orientation. This is corroborated by an isotropic magnetic susceptibility above the Néel temperature [34], and the low value of single-ion anisotropy of $Mn^{2+}$ reported from electron spin resonance measurements [20,35]. As a consequence, perturbations of the crystal field are largely decoupled from the spins. Instead, the absorbed energy is likely dissipated into lower frequency phonons that heat the lattice via anharmonic decay and radiative damping. As the heat capacity of the crystal lattice is generally much higher than that of the spin system, the lattice acts akin to a bath whose temperature is promptly increased, in turn melting the magnetic order [36]. These conclusions are also in line with recent experiments on yttrium iron garnet, where resonant phonon excitation also leads to partial quenching of the ferrimagnetic order [37]. The long recovery time of the melted state could be explained by a low value of interplane heat conduction of $MnPS_3$, typical for layered van der Waals crystals [38]. We remark that excitation of lower energy Raman phonons, which specifically target atomic motions within the superexchange Mn-S-Mn bonds, might couple more strongly to the magnetic order. While these modes could be activated through rectification of the optical phonon [11,39], we see no evidence of this occuring in our system. The weak structural correlation of the infrared and Raman phonons in the thiophosphate family probably excludes their coupling [32], so that electronic Raman scattering might be a more feasible way to excite the low-energy phonons.

In summary, we consider the antiferromagnet $MnPS_3$, whose electrons have zero orbital angular momentum, and drive distinct sub-systems (magnetic $Mn^{2+}$ ions and crystal lattice) out of equilibrium while probing the time-evolution of the magnetic system. Excitation of a singlet—triplet orbital transition of the $Mn^{2+}$ electrons activates the zone-center magnon, whereas lattice excitation leads to quenching of spin order. We argue that the orbital singlet $Mn^{2+}$ ground state prohibits spin-lattice coupling, but that by bringing the magnetic ion out of the singlet ground state, the electrons can interact with the lattice and experience a transient magnetic anisotropy, evidenced by coherent spin evolution. Our results indicate that in insulating magnets with highly spherical magnetic ions, e.g. $Mn^{2+}$ and $Fe^{3+}$, orbital transitions are the more viable paths to achieve resonant optomagnetic control. Other in-gap states, such as spin-coupled excitons, have been fruitful as excitation targets for spin dynamics in other systems [40,41], and should also be considered in $MnPS_3$ [42].

This work was supported by the EU through the European Research Council, grants no. 677458 (AlterMateria) and 788222 (MOL-2D), and by the COST action MOLSPIN CA15128, the Netherlands Organisation for Scientific Research (NWO/OCW) as part of the Frontiers of Nanoscience program (NanoFront), and VENI-VIDI-VICI program, the Spanish MICINN (project MAT-2017-89993-R and Unit of Excellence "Maria de Maeztu" CEX2019-000919-M). M.Š. and H.S.J.v.d.Z. acknowledge funding from the EU Horizon 2020 research and innovation program under grant agreement numbers 785219 and 881603. B.A.I. is supported in part by the National Research Fund of Ukraine within the competition program "Support for research of leading and young scientists," and the project "Development of the physical basis of magnetic nanoelectronics" no. 2020.02/0261.